\documentclass[a4paper]{jpconf}
\usepackage{graphicx}
\begin{document}
\title{XMASS}

\author{Hiroyuki Sekiya, for the XMASS collaboration}

\address{Kamioka Observatory, ICRR, University of Tokyo, 456 Higashi-Mozumi, Kamioka, Hida , Gifu, 506-1205, Japan}

\ead{sekiya@icrr.u-tokyo.ac.jp}

\begin{abstract}
The XMASS detector is a large single phase liquid Xenon scintillator.
After its feasibility had been studied using a 100 kg size prototype 
detector~\cite{minamino}, an 800 kg size detector is being built for 
dark matter search with the sensitivity of $10^{-45} {\rm cm}^2$ region 
in spin-independent cross section. 
The results of R\&D study for 800 kg detector, especially ultra low background 
technologies, and the prospects of the experiment are described.
\end{abstract}

\section{Introduction}
Recently, many dark matter search experiments are underway using liquid Xe
because of its special features~\cite{xereview}.
As sharing many features of liquid Xe with lower cost, liquid Ar is also 
interested for dark matter search~\cite{these}, 
however Xe still has the advantage of density and mass number $A$.

Simultaneous detection of both scintillation of ionization in Xe provides
the nuclear recoil discrimination power, consequently, most experiments
adopted the two-phase time projection detectors.
In contrast, XMASS chose a single-phase scintillator, because
in principle simple detector setups should easily achieve the high radio-purity 
that is mandatory for dark matter search.    
Accordingly, the background suppression technologies should be more critical.

\begin{figure}[ht]
\begin{center}
\includegraphics[width=6.5cm, angle=0, clip]{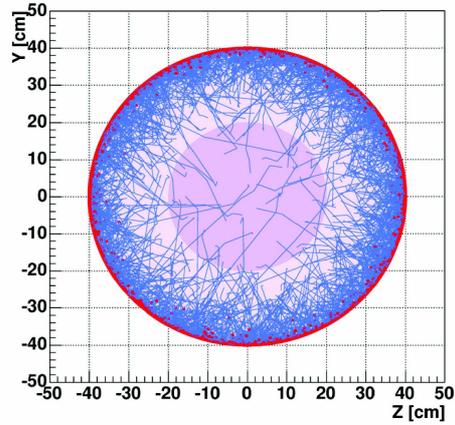}\hspace{2pc}%
\caption{\label{selfshield}Simulation of $^{238}$U-chain gamma rays injected on the surface of the liquid xenon volume. The tracks show the trajectory of gamma rays. The diameter of the liquid xenon volume is 80 cm.}
\end{center}
\end{figure}

The key idea of the background reduction in XMASS is ``self-shielding''. 
The external gamma rays can be absorbed by the liquid xenon itself. 
Figure~\ref{selfshield}. shows a simulation of $^{238}$U-chain
gamma rays injected on the surface of a liquid xenon volume. The outer
surface of the liquid xenon volume absorbs the external gamma rays within
10-20 cm of thickness, and a low background environment in the central 
volume is achieved, while dark matters interact throughout the detector.
Therefore dark matters can be observed in a low background 
environment by extracting the events in the central fiducial volume.

%

\section{The 800kg detector}

\begin{figure}[ht]
\begin{center}
\hspace{-2cm}
\includegraphics[width=12cm, angle=0]{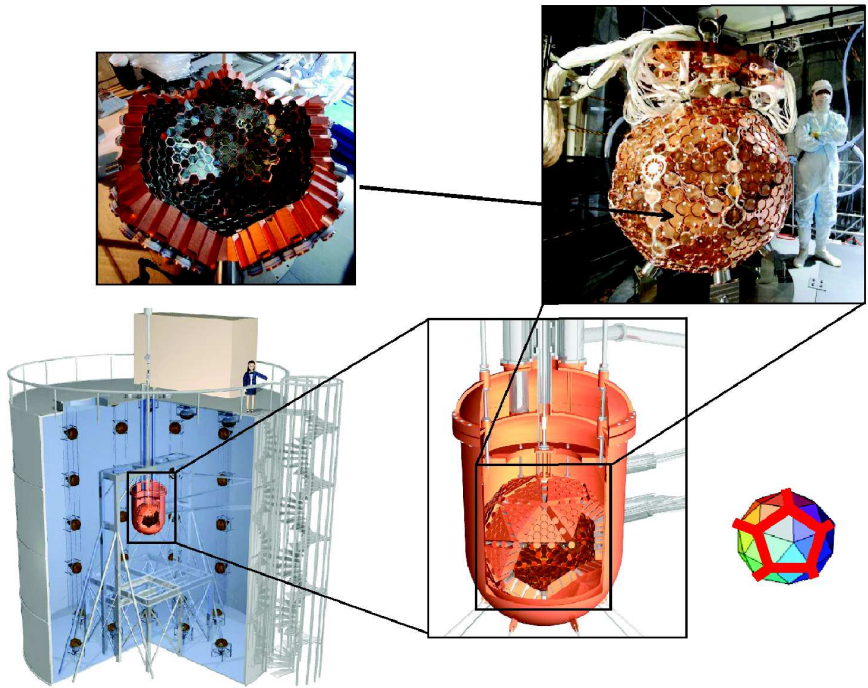}
\caption{\label{800kg}The 800 kg detector.}
\end{center}
\end{figure}

The 800 kg detector, currently being built, is the first designed XMASS
 detector dedicated to search for dark matter.
Actually, the detector consists of 857 kg liquid Xe
in a OFHC vessel, newly developed 642 PMTs, and a 10 m diameter and 10 m
height water tank. 
Figure~\ref{800kg}. shows the schematic drawings of the detector and the 
pictures of the completed ``PMT ball'' for Xe scintillation light readouts.
The PMTs are mounted on a 80 cm diameter pentakis-dodecahedron with 64.2\% 
photocathode coverage of the inner surface. 
This configuration provide a light yield of 4.4 photoelectrons/keV
and the energy threshold below 5 keV (electron equivalent).

\section{Ultra low background technologies}
\subsection{Water tank}

In order to reduce the environmental radioactive background, the 800 kg detector is
put in a water shield. Simulations of gamma rays and fast neutrons show that about a
200 cm thick water layer is enough to reduce the environmental backgrounds to a level
below that of the PMT induced backgrounds, however a water tank 10 m diameter and 10 m
height is prepared not only for this 800 kg detector but also for future projects. 
As shown in Figure~\ref{800kg}., 72 PMTs (20 inch PMT) are attached to the wall of the 
tank to detect the Cherenkov light and thus the water tank can also work as the 
shield for muon induced events.

The water in the tank is circulated and purified with the flow rate of 5 ton/hour by
a dedicated water purification system. The Rn concentration in the water is kept
below 1 mBq/m$^3$.

\subsection{Low background PMTs}

The most serious external backgrounds come from the radioactive contamination in 
the PMTs.
Since 2000, we have developed high radio-pure PMTs with with Hamamatsu 
Photonics K. K. All the components of PMTs were examined individually and 
selected by ourselves. In 2002, R8778, a 2 inch diameter PMT for the 100 kg 
prototype XMASS detector, was developed and it is also used 
in LUX~\cite{xereview} now.
Then, in 2009, we have  succeeded in reducing about a factor 10 of radioactivity
 and developed a hexagonal R10789. 
The PMTs have rather high quantum efficiency, up to  39\% for 178 nm photons at 173 K.

Table~\ref{pmtimp}. shows the contamination of the radioactive impurities 
of R8778 and R10789  measured by a high purity Ge (HPGe) detector in Kamioka underground
laboratory.

\begin{center}
\begin{table}[h]
\caption{\label{pmtimp}Radioactive impurities in the developed PMTs measured by HPGe detector in Kamioka. The uranium and thorium chains were assumed
to be in radioactive equilibrium.}
\centering
\begin{tabular}{@{}*{7}{l}{l}}
\br
RI & R8778 & R10789\\
\mr
$^{238}$U(mBq/PMT) & 18$\pm$2 & 0.70$\pm$0.28\\
$^{232}$Th(mBq/PMT)& 6.9$\pm$1.3 & 1.51$\pm$0.31\\
$^{40}$K(mBq/PMT)& 140$\pm$20 & $<$5.1\\
$^{60}$Co(mBq/PMT)& 18$\pm$2 & 2.92$\pm$1.61\\
\br
\end{tabular}
\end{table}
\end{center}

\subsection{Materials screening}

Not only the components of the PMT, but other components are also screened 
with HPGe measurement. Although our detector is quite simple, more than 250 materials 
are used  around the liquid Xe region, such as OFHC PMT holder, screws, cables, PTFE gaskets, 
LEDs for calibration, and so on.
Sum of radioactive impurities of the measured materials (upper limits) are
listed in Table\ref{otherimp}.

\begin{center}
\begin{table}[h]
\caption{\label{otherimp} Sum of the measured radioactive impurities in the detector components compared with those in 642 PMTs}
\centering
\begin{tabular}{@{}*{7}{c}}
\br
RI & compared with 642 PMTs\\
\mr
$^{238}$U(/PMT) & $<35$\%\\
$^{232}$Th(/PMT)& $<35$\%\\
$^{40}$K(/PMT)& $<20$\%\\
$^{60}$Co(/PMT)& $<20$\%\\
\br
\end{tabular}
\end{table}
\end{center}

\subsection{Kr removal}

\begin{figure}[ht]
\begin{center}
\includegraphics[width=12cm, angle=0]{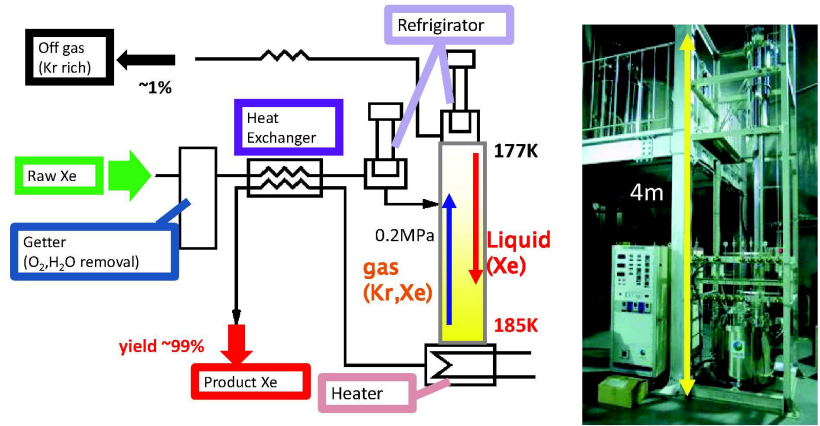}
\vspace{0.8cm}
\caption{\label{distillation}The distillation system}
\end{center}
\end{figure}

As for the internal background, $^{85}$Kr in xenon is one potential 
source due to the $\beta$-decay of  $^{85}$Kr ($\tau=$10.8y, $Q_\beta=$ 687 keV). 
The natural abundance $^{85}$ Kr/Kr  is 1.2$\times$10$^{-10}$ and commercial 
available ``pure Xe'' contains  about 0.1 ppm level of Kr, as a consequence, 
to keep the background from $^{85}$Kr below 10$^{-5}$ counts/day/keV/kg, 
Kr concentration should be below 1ppt, i.e 5 order reduction is required.
In order to achieve this, we have developed new distillation system based 
on our previous system~\cite{distillation}. The distillation technique 
efficiently removes Kr from Xe because the boiling temperature 
of Xe and Kr are quite different(178 K for Xe and around 145 K for 
Kr at the operation pressure of 0.2 atmosphere). The schematic diagram 
and the picture of the system is shown in Figure~\ref{distillation}. 
The system has 99\%  efficiency for collecting xenon (i.e. only 1\% of 
original xenon is rejected) with the process speed 4.7 kg/hour.

\subsection{Rn removal}

\begin{figure}[ht]
\begin{center}
\includegraphics[width=12cm, angle=0]{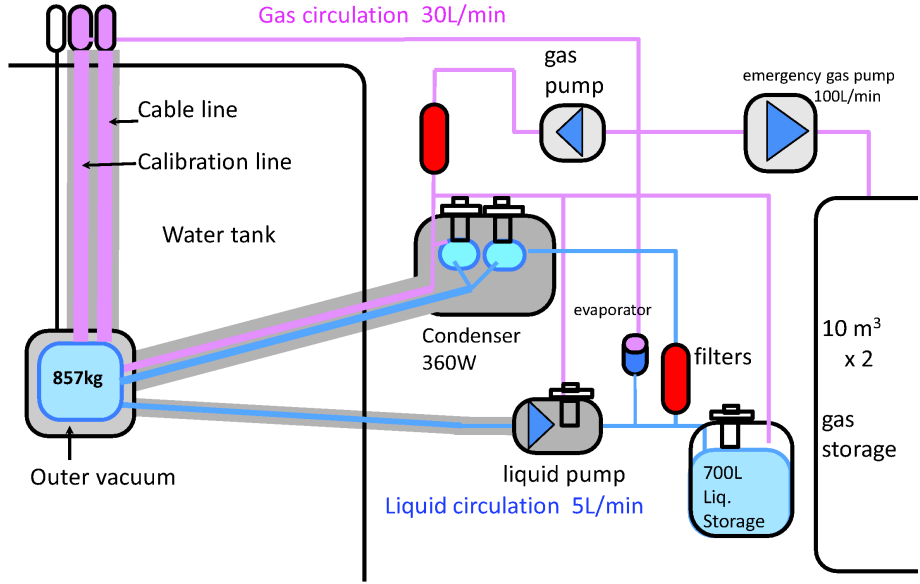}
\vspace{0.7cm}
\caption{\label{circulation}The Xe circulation and purification  system}
\end{center}
\end{figure}

$^{222}$Rn is the most difficult background source because 
$\beta$-decay of $^{222}$Rn daughters remain in Xe; moreover,   
$^{222}$Rn is emanated from all the materials.
During the materials screening, $^{222}$Rn emanation rates
from the components were also measured using high sensitivity Rn detectors
developed for Super-Kamiokande~\cite{rn}. It is found that up to $15$ mBq
sources exist in the detector, thus these $^{222}$Rn should be continuously removed 
from the system. Figure\ref{circulation} shows the purification system. 

In order to suppress the background from $^{222}$Rn as the level of those from the PMTs,
Rn concentration should be kept below 1.2 mBq/ton in liquid phase 
or 0.8 $\mu$Bq/m$^3$ in gas phase. 
We have succeeded removing $^{222}$Rn from gas phase using 
cooled ($-100^{\circ}$C) activated charcoals with more than 90\%  efficiency, 
however as the flow rate
of gas circulation is limited by the cooling power of the condenser, 
Rn removal from liquid is preferable.  We are testing some setups in liquid phase now.

\section{Expected sensitivity}
\begin{figure}[ht]
\begin{center}
\includegraphics[width=5.5cm]{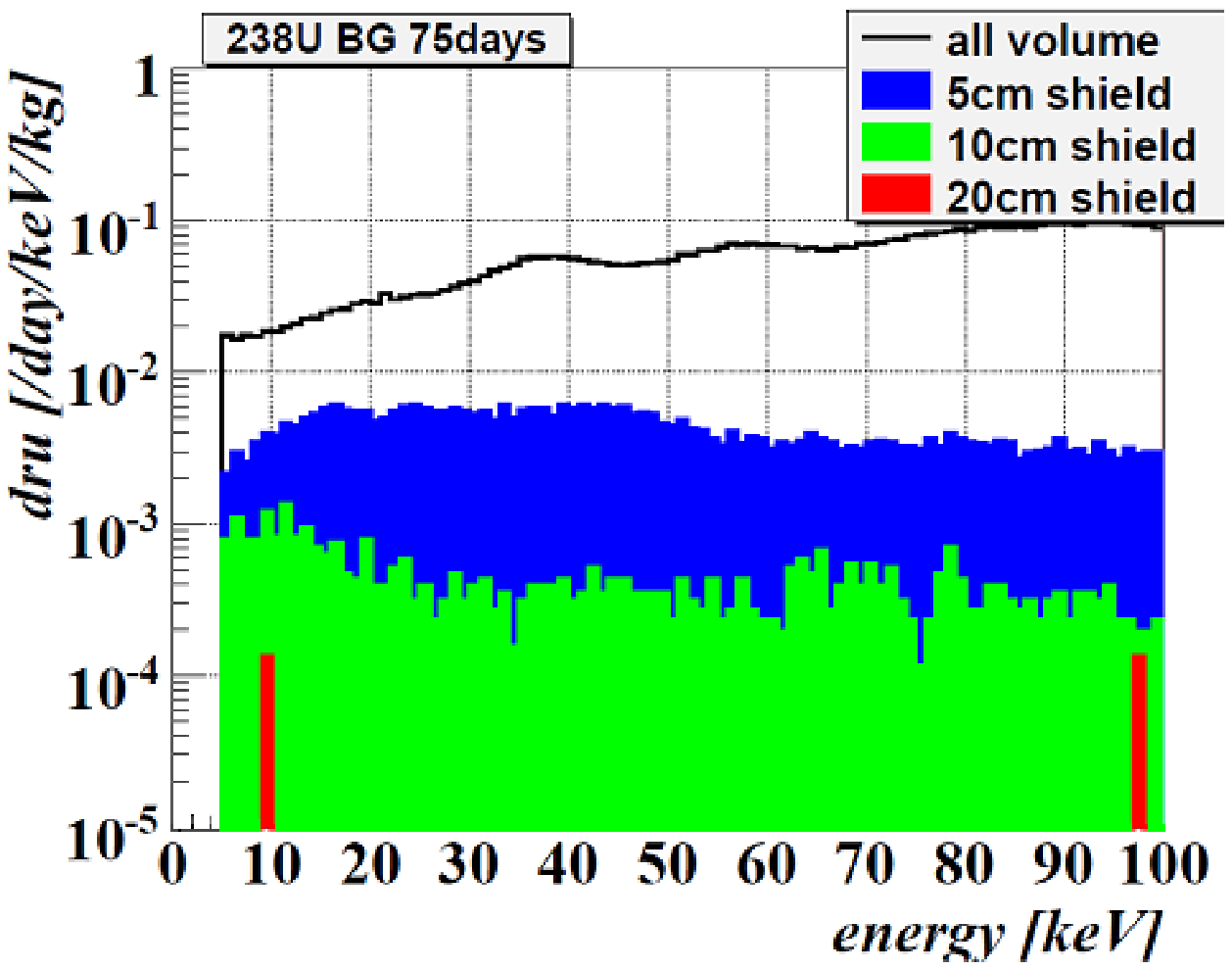}
\includegraphics[width=6cm]{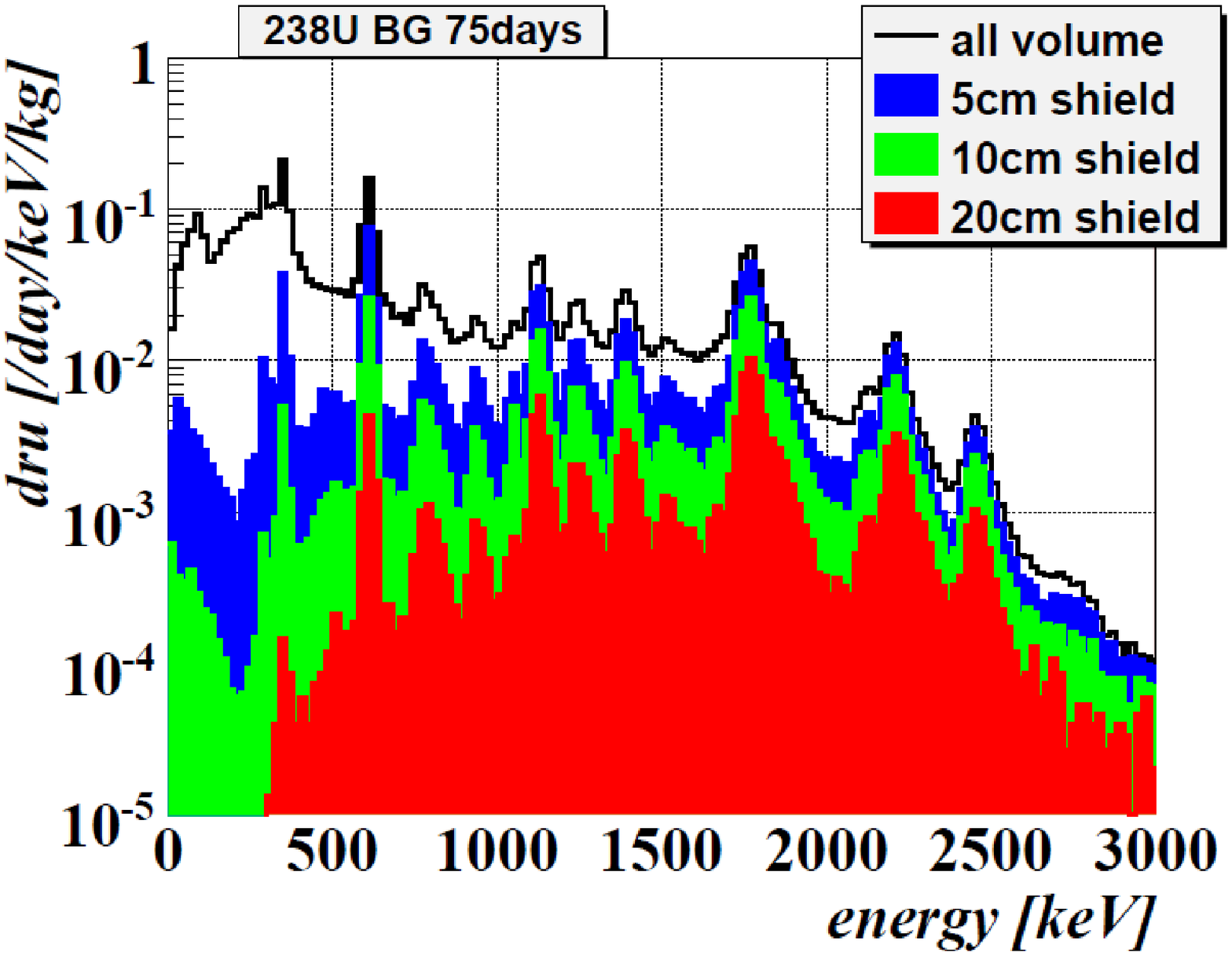}
\caption{\label{upmt}The evaluated background rate from $^{238}$U in 642 PMTs. 20 cm of the self-shield layer
corresponds to the fiducial volume of 100 kg.}
\end{center}
\end{figure}

By putting above listed background sources into the full detector MC simulation
code, the total backgrounds in fiducial volume of 100 kg was evaluated and
turned out to be below 10$^4$ counts/day/kg/keVee.  
As an example and in order to demonstrate the ``self-shielding'' effect, 
the background contribution from  $^{238}$U-chain in PMTs is shown 
in Figure~\ref{upmt}.

\begin{figure}[ht]
\begin{center}
\includegraphics[width=7cm]{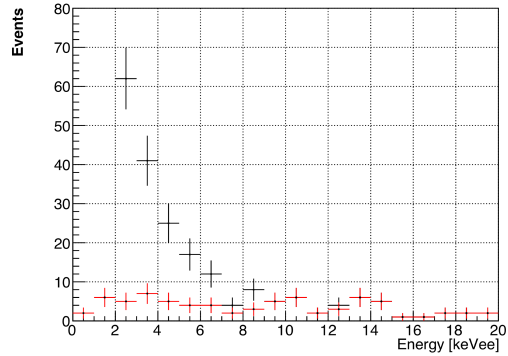}
\caption{\label{energy}The expected energy spectrum in the 100 kg fiducial volume. Red points show the
background events and black points show the background $+$ the expected dark matter signal when M$_\chi=50$ GeV and $\sigma_{\chi\rm{-p}}=3\times10^{-44}$ cm$^{-2}$.}
\end{center}
\end{figure}

\begin{figure}[ht]
\begin{center}
\includegraphics[width=7cm]{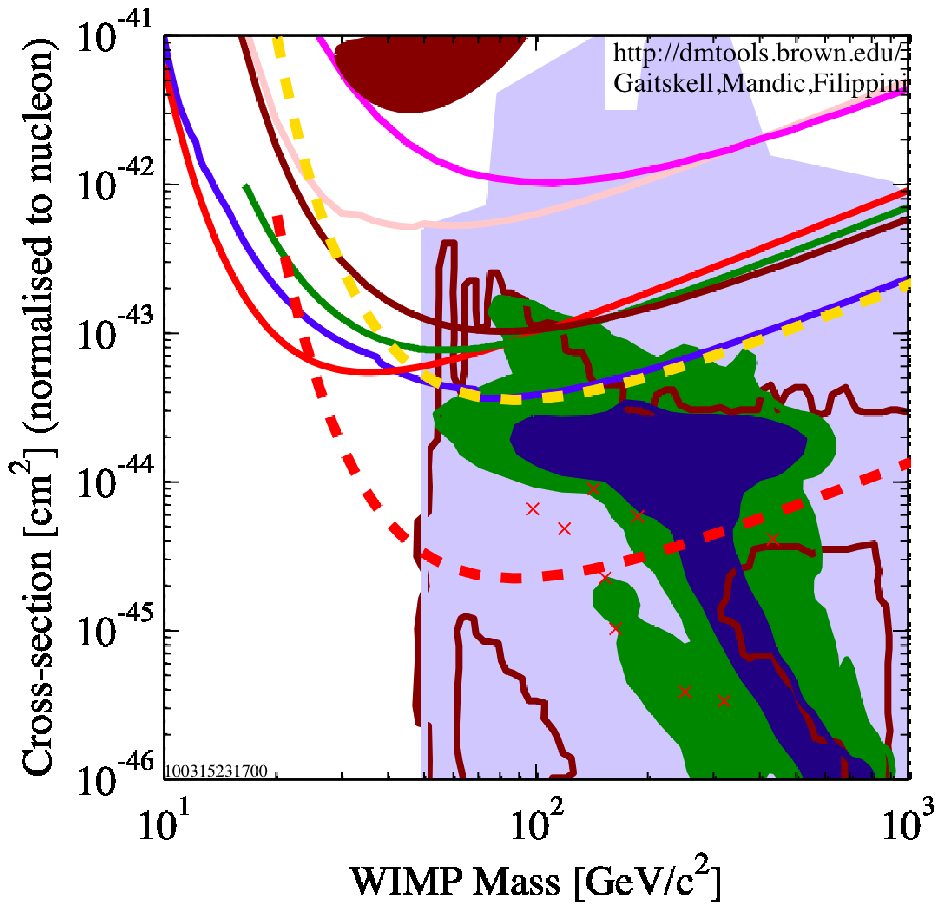}
\includegraphics[width=5cm]{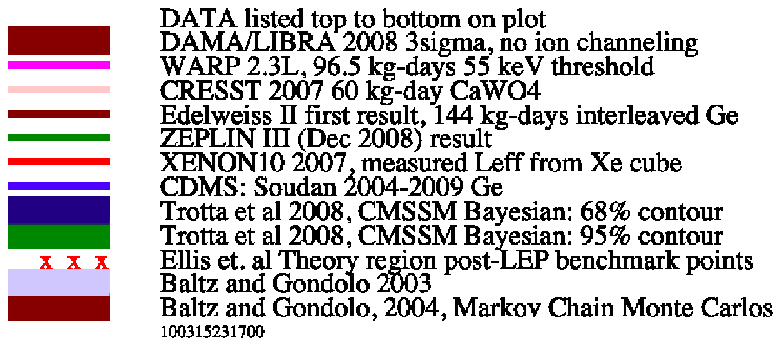}
\caption{\label{sens}The expected sensitivity of the 800kg detector compared with other
experiments. The yellow dotted is the 90\% CL spin-independent sensitivity line for 10 days operation and the red dotted is the one for 1 year operation}
\end{center}
\end{figure}

With these background assumptions, the expected detection sensitivity is derived.
Figure~\ref{energy} shows the expected energy spectrum in case of 
M$_\chi$=50GeV and $\sigma_{\chi\rm{-p}}=3\times10^{-44}$cm$^{-2}$,
and 90\% CL spin-independent sensitivity lines for 10 days operation and 
1 year operation are shown in Figure~\ref{sens}.

\section{Conclusion}
The construction of XMASS 800 kg detector is almost finished in Kamioka underground 
laboratory. Its sensitivity for spin-independent interaction of M$_\chi=50$ GeV case 
is expected to be $\sigma_{\chi\rm{-p}}=2\times10^{-44}$ cm$^{-2}$
with one year operation.
After the commissioning runs for a few months, the dark matter search run will be started 
in this year, 2010.

\end{document}